\documentclass{PoS}

\newcommand{\gsim}{\mbox{ \raisebox{-1.0ex}{$\stackrel{\textstyle >}
{\textstyle \sim}$ }}}
\newcommand{\lsim}{\mbox{ \raisebox{-1.0ex}{$\stackrel{\textstyle <}
{\textstyle \sim}$ }}}

\title{A TeV-scale model for neutrino mass, DM and baryon asymmetry  }

\ShortTitle{A TeV-scale model for neutrino mass, DM and baryon asymmetry}

\author{Mayumi Aoki\\
        Department of Physics, Tohoku University, Aramaki, Aoba,
        Sendai, Miyagi 980-8578, JAPAN \\
        E-mail: \email{mayumi@tuhep.phys.tohoku-u.ac.jp}}

\author{\speaker{Shinya Kanemura}\\ 
        Department of Physics, Univeristy of Toyama, 3190 Gofuku, Toyama 930-8555, JAPAN\\
        E-mail: \email{kanemu@sci.u-toyama.ac.jp}}

\author{Osamu Seto\\
        Instituto~de~F\'{i}sica~Te\'{o}rica~UAM/CSIC,~Universidad~Aut\'{o}noma~de~Madrid,
        Cantoblanco,
        Madrid 28049, Spain\\
        E-mail: \email{osamu.seto@uam.es}}

\abstract{
We discuss a model which would explain neutrino oscillation, dark matter, 
and baryon asymmetry of the Universe simultaneously by the physics at TeV 
scales. Tiny neutrino masses are generated at the three loop level due to 
the exact $Z_2$ symmetry, by which the stability of the dark matter candidate 
is also guaranteed. The extra Higgs doublet is required not only for the 
tiny neutrino masses but also for successful electroweak baryogenesis.
The model provides various discriminative predictions especially in
charged Higgs 
phenomenology.  }

\FullConference{Prospects for Charged Higgs Discovery at Colliders\\
		 September 16-19 2008\\
		 Uppsala, Sweden}

\begin{document}

\section{Introduction} 

Today, we have clear motivations to consider a model beyond the standard model (SM).
The data indicate that neutrinos have tiny masses and mix with each other\cite{lep-data}. 
Cosmological observations have revealed that the energy density of dark
matter (DM) in the Universe dominates that of baryonic matter\cite{wimp}.
Asymmetry of matter and anti-matter in our Universe has been addressed
for several decades\cite{sakharov}.

In this talk, we discuss a model which would explain neutrino
oscillation, origin of DM and baryon asymmetry simultaneously 
by an extended Higgs sector with TeV-scale right-handed (RH) neutrinos\cite{aks}. 
Tiny neutrino masses are generated at the three loop level due to an
exact discrete symmetry, by which tree-level Yukawa couplings of neutrinos are prohibited.
The lightest neutral odd state under the discrete symmetry is a candidate of DM. 
Baryon asymmetry can be generated at the electroweak phase transition
(EWPT) by additional CP violating phases in the Higgs sector\cite{ewbg-thdm}.
In this framework, a successful model can be built without contradiction
of the current data.

Original idea of generating tiny neutrino masses via the radiative effect 
has been proposed by Zee\cite{zee}. 
The extension with a TeV-scale  RH neutrino has been discussed in Ref.~\cite{knt},
where neutrino masses are generated at the three-loop level due to the exact $Z_2$
parity, and the $Z_2$-odd RH neutrino is a candidate of DM. This 
has been extended with two RH neutrinos to describe the neutrino data\cite{kingman-seto}. 
Several models with adding baryogenesis have been considered in Ref.~\cite{ma}.
The following advantages would be in the present model:
(a)~all mass scales are at most at the TeV scale without large hierarchy, 
(b)~physics for generating neutrino masses is connected with that for
DM and baryogenesis, 
(c)~the model parameters are strongly constrained by the current data, so
    that the model provides discriminative predictions which can be tested
    at future experiments.
  
\section{Model}

We introduce two scalar isospin doublets with hypercharge $1/2$ ($\Phi_1$ and $\Phi_2$),  
charged singlets ($S^\pm$), a real scalar singlet ($\eta$) and two
generation isospin-singlet RH neutrinos ($N_R^\alpha$ with $\alpha=1, 2$).
As in Ref.~\cite{knt} we impose an exact $Z_2$ symmetry to generate tiny
neutrino masses at the three-loop level, which we refer as $Z_2$. 
We assign $Z_2$-odd charge to $S^\pm$, $\eta$ and $N_R^\alpha$, while 
ordinary gauge fields, quarks and leptons and Higgs doublets are  $Z_2$ even.
Introduction of two Higgs doublets would cause a dangerous flavor
changing neutral current. To avoid this in a natural way, we impose
another discrete symmetry  ($\tilde{Z}_2$) that is softly broken\cite{glashow-weinberg}.
From a phenomenological reason discussed later, we assign $\tilde{Z}_2$ charges such that
only $\Phi_1$ couples to leptons whereas $\Phi_2$ does to quarks;  
\begin{eqnarray}
 {\cal L}_Y  \!=\! -\!  y_{e_i}^{}  \overline{L}^i \Phi_1 e_R^i
     \! - \! y_{u_i}^{}  \overline{Q}^i \tilde{\Phi}_2 u_R^i
     \! - \! y_{d_i}^{}  \overline{Q}^i \Phi_2 d_R^i + {\rm h.c.}, \label{typex-yukawa}
\end{eqnarray}
where $Q^i$ ($L^i$) is the ordinary $i$-th generation left-handed (LH) quark (lepton)
doublet,  and $u_R^i$ and $d_R^i$ ($e_R^i$) are RH-singlet up- and
down-type quarks (charged leptons), respectively.   
We summarize the particle properties under 
$Z_2$ and $\tilde{Z}_2$ in Table~\ref{discrete}.
\begin{table}[t]
\begin{center}
  \begin{tabular}{c|ccccc|cc|ccc}
   \hline
   & $Q^i$ & $u_R^{i}$ & $d_R^{i}$ & $L^i$ & $e_R^i$ & $\Phi_1$ & $\Phi_2$ & $S^\pm$ &
    $\eta$ & $N_{R}^{\alpha}$ \\\hline
$Z_2\frac{}{}$                ({\rm exact}) & $+$ & $+$ & $+$ & $+$ & $+$ & $+$ & $+$ & $-$ & $-$ & $-$ \\ \hline  
$\tilde{Z}_2\frac{}{}$ ({\rm softly\hspace{1mm}broken})& $+$ & $-$ & $-$ & $+$ &
                       $+$ & $+$ & $-$ & $+$ & $-$ & $+$ \\\hline
   \end{tabular}
\end{center}
  \caption{Particle properties under the discrete symmetries.
 }
  \label{discrete}
 \end{table}
Notice that the Yukawa coupling in Eq.~(\ref{typex-yukawa})
is different from that in the minimal supersymmetric SM (MSSM)\cite{hhg}.
In addition to the usual potenital of the two Higgs doublet model (THDM) with
the $\tilde{Z}_2$ parity and that of  the $Z_2$-odd scalars,
we have the interaction terms between $Z_2$-even and -odd scalars:  
\begin{eqnarray}
{\cal L}_{int} = -
 \sum_{a=1}^2 \left(\rho_a |\Phi_a|^2|S|^2 + \sigma_a |\Phi_a|^2
  \frac{\eta^2}{2}\right)
-\sum_{a,b=1}^2\left\{ \kappa \,\,\epsilon_{ab} (\Phi^c_a)^\dagger
                    \Phi_b S^- \eta + {\rm h.c.}\right\},
 \end{eqnarray}
where $\epsilon_{ab}$ is the anti-symmetric tensor with $\epsilon_{12}=1$.
The mass term and the interaction for $N_R^\alpha$ are given by 
\begin{eqnarray}
 {\cal L}_{Y_N^{}} \!= \! \sum_{\alpha=1}^2\!\left\{ \!\frac{1}{2}m_{N_R^\alpha}^{} \overline{{N_R^\alpha}^c} N_R^\alpha
                 -  h_i^\alpha \overline{(e_R^i)^c}
                   N_R^\alpha S^-\! + {\rm h.c.}\!\right\}.
\end{eqnarray} 
Although the CP violating phase in the Lagrangian is
crucial for successful baryogenesis at the EWPT\cite{ewbg-thdm},
it does not much affect the following discussions. Thus, we neglect it for simplicity.
We later give a comment on the case with the non-zero CP-violating phase. 

As $Z_2$ is exact, the even and odd fields cannot mix.
Mass matrices for the $Z_2$ even scalars are diagonalized as in the
usual THDM 
by the mixing angles $\alpha$ and $\beta$, where $\alpha$
diagonalizes the CP-even states, and
$\tan\beta=\langle \Phi_2^0
\rangle/\langle \Phi_1^0 \rangle$\cite{hhg}. 
The $Z_2$ even physical states are two CP-even ($h$ and $H$),
a CP-odd ($A$) and charged ($H^\pm$) states.
We here define $h$ and $H$ such that $h$ is always
the SM-like Higgs boson when $\sin(\beta-\alpha)=1$. 

\section{Neutrino Mass, Dark Matter, and Strongly 1st-Order Phase Transition}

\begin{figure}
\begin{center}
 \includegraphics[width=.6\textwidth]{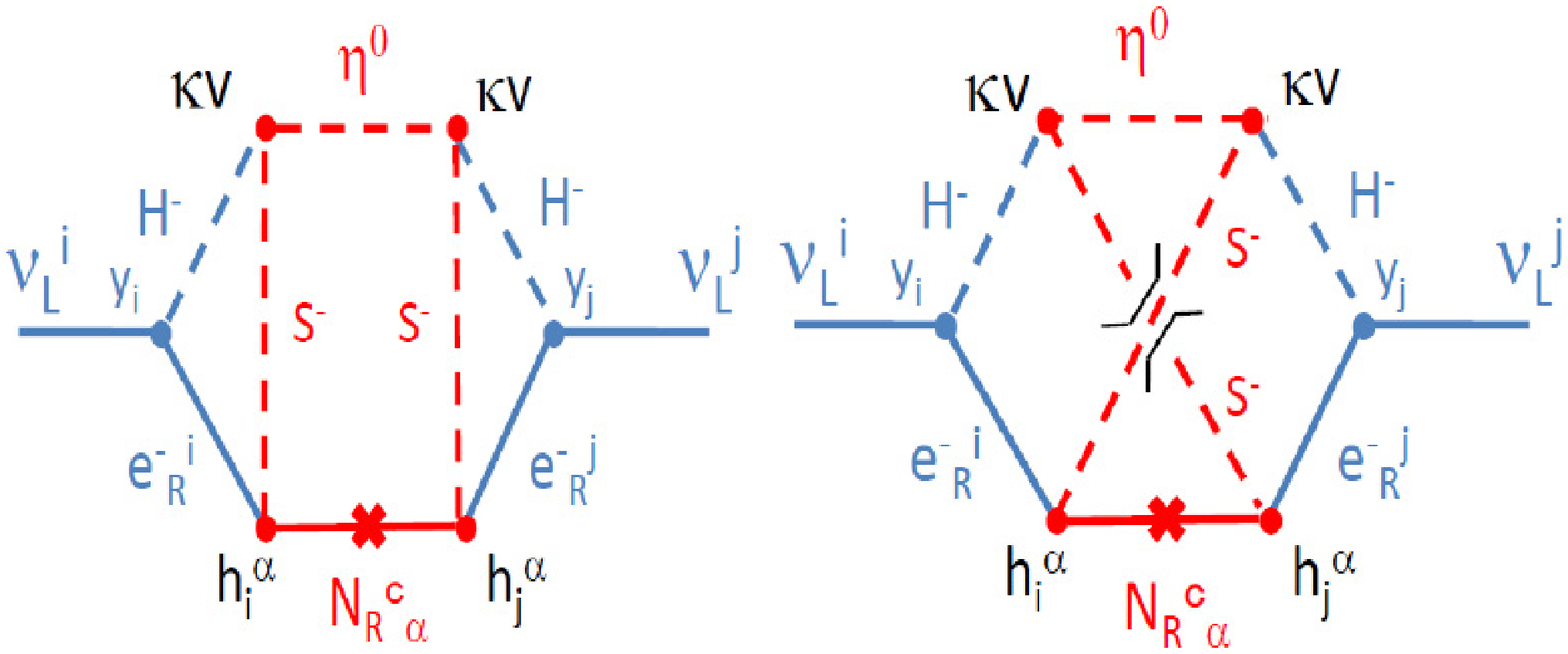}
  \caption{The diagrams for generating tiny neutrino masses. }
  \label{diag-numass}
\end{center}
\end{figure}

The LH neutrino mass matrix $M_{ij}$ is generated by the three-loop diagrams in Fig.~\ref{diag-numass}.
The absence of lower order loop contributions is guaranteed by $Z_2$.
$H^\pm$  and  $e_R^i$ play a crucial role to connect LH neutrinos with the one-loop sub-diagram
by the $Z_2$-odd states.
We obtain
\begin{eqnarray}
M_{ij} = \sum_{\alpha=1}^{2} 
  C_{ij}^\alpha F(m_{H^{\pm}}^{},m_{S^{\pm}}^{},m_{N_R^{\alpha}}^{}, m_\eta), 
\end{eqnarray}
where $C_{ij}^\alpha =
   4 \kappa^2 \tan^2\!\beta 
  (y_{e_i}^{\rm SM} h_i^\alpha) (y_{e_j}^{\rm SM} h_j^\alpha)$ with 
$y_{e_i}^{\rm SM}=\sqrt{2}m_{e_i}/v$ and  $v\simeq 246$GeV.
  The factor of the three-loop integral function
   $F(m_{H^{\pm}}^{},m_{S^{\pm}}^{},m_{N_R}^{}, m_\eta)$
  includes the supression factor of $1/(16\pi^2)^3$, whose typical size
  is ${\cal O}(10^{4})$eV.
Magnitudes of $\kappa \tan\beta$ as well as $F$
determine the universal scale of $M_{ij}$, 
whereas variation of $h_i^\alpha$ ($i=e$, $\mu$, $\tau$) 
reproduces the mixing pattern indicated by the neutrino
 data\cite{lep-data}.
%
%
\begin{table}
\begin{center}
  \begin{tabular}{|c||c|c|c|c|c|c|c|}\hline
     Set   & $h_e^1$ & $h_e^2$ & $h_\mu^1$ & $h_\mu^2$ & $h_\tau^1$ & $h_\tau^2$  &
   $B(\mu\!\!\to\!\! e\gamma)$ \\\hline 
  A &  2.0    &  2.0     &  -0.019     & 0.042
                   &-0.0025   & 0.0012  & $6.9\!\times \!10^{-12}$  \\\hline 
   B & 2.2     &  2.2     &  0.0085     & 0.038 
                   & -0.0012  &    0.0021  & $6.1\!\times \!10^{-12}$ \\\hline 
   \end{tabular}
\end{center}
 \caption{Values of $h_i^\alpha$ for $m_{H^\pm}^{} (m_{S^\pm}^{})=100(400)$GeV 
  $m_\eta=50$ GeV, $m_{N_R^1}=m_{N_R^2}=$3.0 TeV for the normal hierarchy. For  Set A (B), 
  $\kappa\tan\beta=28 (32)$ and $U_{e3}=0 (0.18)$. 
 Predictions on the branching ratio of $\mu\to e
 \gamma$ are also shown.}
  \label{h-numass}
 \end{table}

 Under the {\it natural} requirement 
$h_e^\alpha \sim {\cal O}(1)$, and taking 
the  $\mu\to e\gamma$ search results into account\cite{lfv-data},   
we find that $m_{N_R^\alpha}^{} \sim {\cal O}(1)$ TeV, 
$m_{H^\pm}^{} \lsim {\cal O}(100)$ GeV, $\kappa \tan\beta \gsim {\cal
O}(10)$, and $m_{S^\pm}^{}$ being several
times 100 GeV. 
On the other hand, the LEP direct search results indicate 
$m_{H^\pm}^{}$ (and $m_{S^\pm}^{}$)  $\gsim 100$ GeV\cite{lep-data}.  
In addition, with the LEP precision measurement for the $\rho$ parameter,  
possible values uniquely turn out to be  
$m_{H^\pm}^{} \simeq m_{H}^{}$ (or $m_{A}^{}$) $\simeq 100$ GeV
for $\sin(\beta-\alpha) \simeq 1$. 
Thanks to the Yukawa coupling in Eq.~(\ref{typex-yukawa}), such
a light $H^\pm$ is not excluded by the $b \to s \gamma$ data\cite{bsgamma}.
Since we cannot avoid to include the hierarchy among $y_i^{\rm SM}$,  
we only require $h_i^\alpha y_i \sim {\cal O}(y_e) \sim 10^{-5}$ 
for values of $h_i^\alpha$. 
Our model turns out to prefer the normal hierarchy
scenario. 
Several sets for $h_i^\alpha$ are shown in Table~\ref{h-numass} with the
predictions on the branching ratio of $\mu\to e\gamma$ 
assuming the normal hierarchy.
%
%

\begin{figure}
 \includegraphics[width=.5\textwidth]{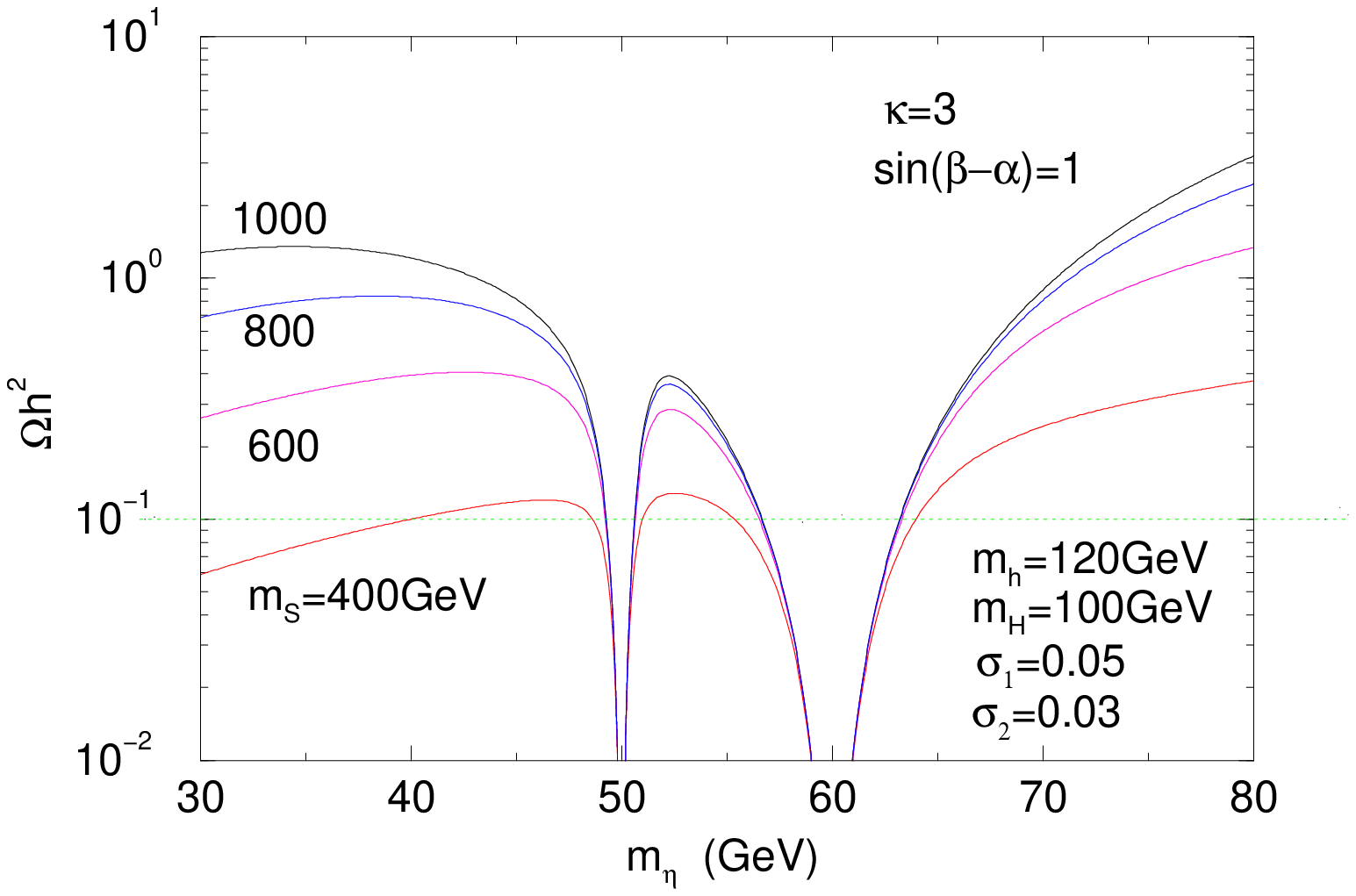}
 \includegraphics[width=.42\textwidth]{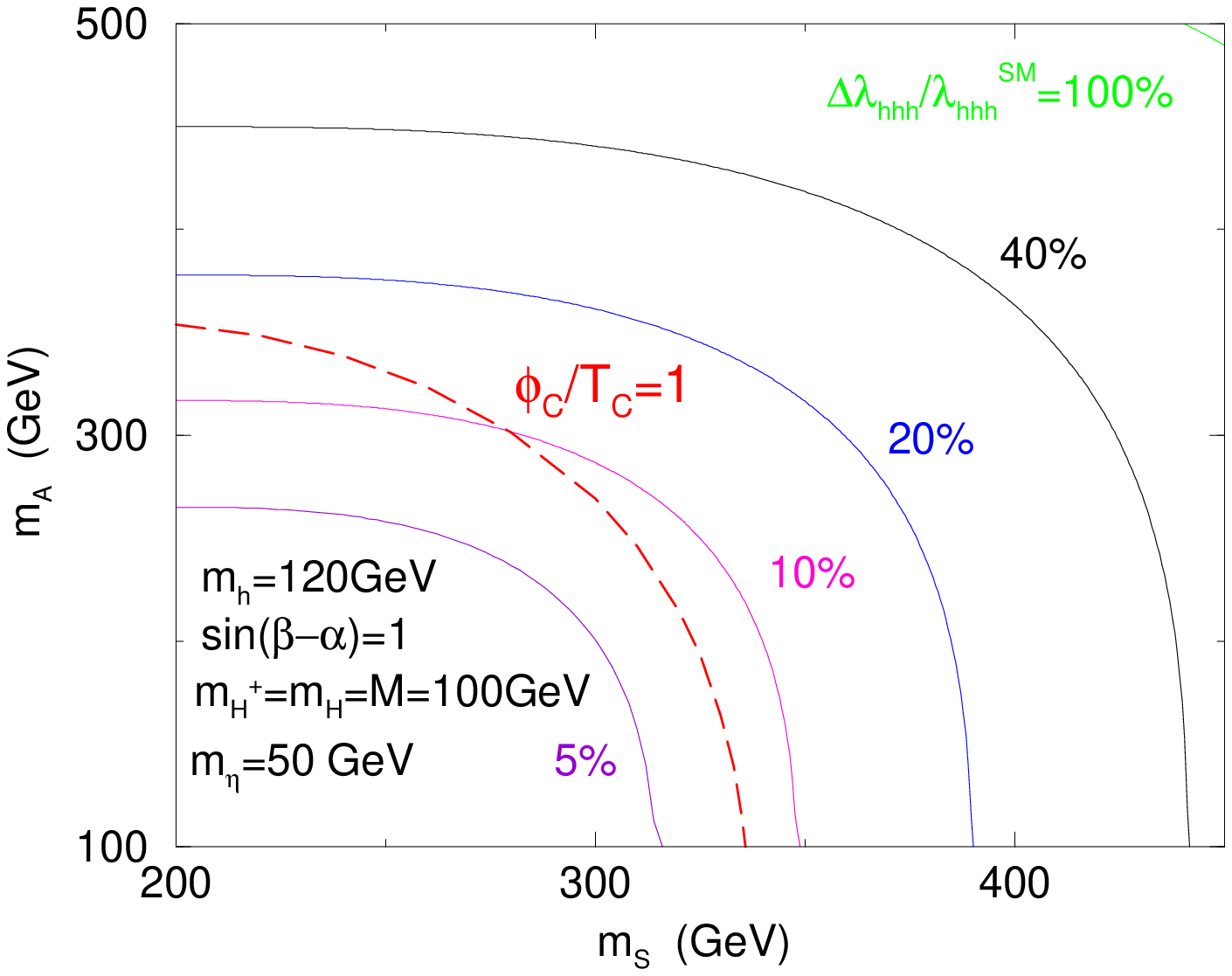}
  \caption{[Left figure] The relic abundance of $\eta$.
 [Right figure] The region of strong first order EWPT.
 Deviations from the SM value in the $hhh$ coupling
 are also shown. }
  \label{etaOmega}
\end{figure}

\indent
The lightest $Z_2$-odd particle is 
stable and can be a candidate of DM if it is neutral.
In our model, $N_R^\alpha$ must be heavy, so that 
the DM candidate is identified as $\eta$.
When $\eta$ is lighter than the W boson, $\eta$ dominantly annihilates 
into $b \bar{b}$ and $\tau^+\tau^-$ via tree-level $s$-channel
Higgs ($h$ and $H$) exchange diagrams, and into $\gamma\gamma$ via
one-loop diagrams.
From their summed thermal averaged annihilation rate $\langle \sigma v \rangle$,
the relic mass density  $\Omega_\eta h^2$ is 
evaluated.
Fig.~\ref{etaOmega}(Left) shows 
$\Omega_{\eta}h^2$ as a function of $m_\eta$. 
Strong annihilation can be seen near $50$ GeV $\simeq m_H^{}/2$
($60$ GeV $\simeq m_h/2$) due to the resonance of $H$ ($h$) mediation.
The data ($\Omega_{\rm DM} h^2 \simeq 0.11$\cite{wimp}) indicate that $m_\eta$ is around 40-65 GeV. 
%


The model satisfies the necessary 
conditions for baryogenesis\cite{sakharov}.
Especially, departure from thermal equilibrium can be 
realized by the strong first order EWPT.
The free energy is given at a high temperature $T$ by
\begin{eqnarray}
 V_{eff}[\varphi, T]= D (T^2-T_0^2) \varphi^2 
                     - E T \varphi^3 
                     + \frac{\lambda_T}{4} \varphi^4 + ..., 
\end{eqnarray}
where $\varphi$ is the order parameter.
A large value of the coefficient $E$
is crucial for the strong first order EWPT with keeping
$m_h \lsim 120$ GeV. 
For sufficient sphaleron decoupling in the broken phase, it is required that\cite{sph-cond} 
\begin{eqnarray}
 \frac{\varphi_c}{T_{c}}  \left(\simeq \frac{2 E}{\lambda_{T_c}}\right) 
   \gsim 1, \label{sph2}
\end{eqnarray}
where $\varphi_c$ ($\neq 0$) and $T_c$ are the critical values of
$\varphi$ and $T$ at the EWPT.
In Fig.~\ref{etaOmega}(Right), the allowed region under the condition of
Eq.~(\ref{sph2}) is shown. The condition is satisfied
when
$m_{S^{\pm}}^{} \gsim 350$ GeV
for $m_A^{} \gsim 100$ GeV, 
$m_h \simeq 120$ GeV, $m_H^{} \simeq m_{H^\pm}^{} (\simeq 
M) \simeq 100$ GeV and $\sin(\beta-\alpha)\simeq 1$.  

\section{Phenomenology}

A successful scenario which can simultaneously solve the above three issues 
under the data\cite{lep-data,lfv-data,bsgamma} would be 
\begin{eqnarray}
 \begin{array}{ll}
 \sin(\beta-\alpha) \simeq 1, &\!\! (\kappa \tan\beta) \simeq 30, \\
 m_h = 120 {\rm GeV},     &\!\! m_H^{} \simeq m_{H^\pm} (\simeq M) \simeq 100 {\rm GeV},    \\
 m_A \gsim 100 {\rm GeV},     &\!\! m_{S^\pm}^{}\sim 400{\rm GeV},\\
 m_{\eta} \simeq 40-65 {\rm GeV}, &\!\! m_{N_R^{1}} \simeq m_{N_R^{2}}
  \simeq 3 {\rm TeV}.\\
  \end{array} \label{scenario}
\end{eqnarray}
This is realized without assuming unnatural hierarchy among the
couplings. All the masses are between
${\cal O}(100)$ GeV and ${\cal O}(1)$ TeV. As they are required by the data, 
the model has a predictive power. We note that the masses of $A$ and $H$
can be exchanged with each other. 

\indent
We outline phenomenological predictions in the scenario in~(\ref{scenario}) in
order.  The detailed analysis is shown elsewhere\cite{aks-full}.
(I)~$h$ is the SM-like Higgs boson, but decays into $\eta\eta$ when $m_\eta < m_h/2$.
        The branching ratio is about 36\% (25\%) for   $m_\eta \simeq 45$ (55) GeV.
        This is related to the DM abundance, so that  
         our DM scenario is testable at the LHC.
(II)~$\eta$  is potentially detectable
        by direct DM searches\cite{xmass},    
         because $\eta$ can scatter with nuclei 
        via the scalar exchange\cite{john}. 
(III)~For successful baryogenesis, the $hhh$ coupling
      has to deviate from the SM value by more 
        than 10-20 \%\cite{ewbg-thdm2} (see Fig.~\ref{etaOmega}), 
        which can be tested at the International Linear Collider (ILC)\cite{hhh-measurement}.
(IV)~$H$ (or $A$) can predominantly decay into $\tau^+\tau^-$ instead of $b\bar b$ for $\tan\beta\gsim 3$. 
        When $A$ (or $H$) is relatively heavy it can decay into $H^\pm W^\mp$ and $H Z$ (or $A Z$). 
(V)~the scenario with light $H^\pm$ and $H$ (or $A$) can be directly tested at the LHC
    via $pp\to W^\ast \to H H^\pm$ and $A H^\pm$\cite{wah}. 
(VI)~$S^\pm$ can  be produced in pair 
        at the LHC (the ILC)\cite{zee-ph}, and decay into
        $\tau^\pm \nu \eta$. 
        The signal would be a hard hadron pair\cite{hagiwara} with a
        large missing energy.
(VII)~The couplings $h_i^\alpha$ cause lepton flavor violation 
     such as $\mu\to e\gamma$ which would provide information
     on $m_{N_R^{\alpha}}$ at future experiments. 

     Finally, we comment on the case with the CP violating phases.
Our model includes the THDM, so that the same discussion can be applied
in evaluation of  baryon number at the EWPT\cite{ewbg-thdm}.  
The mass spectrum  would be changed to some extent, but most of the features
discussed above should be conserved with a little modification.

\section{Summary}

We have discussed the model with the extended Higgs sector and
TeV-scale RH neutrinos, which would explain neutrino mass and mixing, DM and baryon
asymmetry by the TeV scale physics without less fine tuning. It gives
specific predictions on many low energy phenomena. In particular, charged
Higgs phenomenology in this model is completely different from that in
the MSSM, so that it is testable at the LHC and the ILC.


\begin{thebibliography}{99}
\bibitem{lep-data}
        W.~M.~Yao, et al., [Particle Data Group]
        J.~Phys.~G {\bf 33}
        (2006) 1.
 \bibitem{wimp}
         E.~Komatsu, et al., (WMAP Collaboration),
         arXiv:0803.0547 [astro-ph].
\bibitem{sakharov}
  A.~D.~Sakharov,
  Pisma Zh.\ Eksp.\ Teor.\ Fiz.\  {\bf 5}, 32 (1967).
\bibitem{aks}
        M. Aoki, S. Kanemura and O. Seto, arXiv:0807.0361 [hep-ph].
 \bibitem{ewbg-thdm}
  J.~M.~Cline, K.~Kainulainen and A.~P.~Vischer,
  Phys.\ Rev.\  D {\bf 54}, 2451 (1996);
  L.~Fromme, S.~J.~Huber and M.~Seniuch,
  JHEP {\bf 0611}, 038 (2006).
 \bibitem{zee}
  A.~Zee,
  Phys.\ Lett.\  B {\bf 93}, 389 (1980)
  [Erratum-ibid.\  B {\bf 95}, 461 (1980)];
  A.~Zee,
  Phys.\ Lett.\  B {\bf 161}, 141 (1985).
 \bibitem{knt}
  L.~M.~Krauss, S.~Nasri and M.~Trodden,
  Phys.\ Rev.\  D {\bf 67}, 085002 (2003).
 \bibitem{kingman-seto}
  K.~Cheung and O.~Seto,
  Phys.\ Rev.\  D {\bf 69}, 113009 (2004).
 \bibitem{ma}
  E.~Ma,
  Phys.\ Rev.\  D {\bf 73}, 077301 (2006);
  J.~Kubo, E.~Ma and D.~Suematsu,
  Phys.\ Lett.\  B {\bf 642}, 18 (2006);
  T.~Hambye, K.~Kannike, E.~Ma and M.~Raidal,
  Phys.\ Rev.\  D {\bf 75}, 095003 (2007); 
   K.~S.~Babu and E.~Ma,
   arXiv:0708.3790 [hep-ph]; 
   N.~Sahu and U.~Sarkar,
   arXiv:0804.2072 [hep-ph].
\bibitem{glashow-weinberg}
  S.~L.~Glashow and S.~Weinberg,
  Phys.\ Rev.\  D {\bf 15}, 1958 (1977); 
%
  V.~D.~Barger, J.~L.~Hewett and R.~J.~N.~Phillips,
  Phys.\ Rev.\  D {\bf 41}, 3421 (1990).
 \bibitem{hhg}
  J.~F.~Gunion, et al., 
  ``{\it The Higgs Hunters's Guide}'' (Addison Wesley, 1990).
\bibitem{lfv-data}
   A.~Baldini,
   Nucl.\ Phys.\ Proc.\ Suppl.\  {\bf 168}, 334 (2007).
 \bibitem{bsgamma}
   E.~Barberio,  et al.,  [Heavy Flavor Averaging Group (HFAG)],
   arXiv:hep-ex/0603003.
 \bibitem{sph-cond}   
  G.~D.~Moore,
  Phys.\ Lett.\  B {\bf 439}, 357 (1998);
  Phys.\ Rev.\  D {\bf 59}, 014503 (1998).
\bibitem{aks-full}
M.~Aoki, S.~Kanemura and O.~Seto, in preparation.
\bibitem{xmass}
 Y.~D.~Kim,
 Phys.\ Atom.\ Nucl.\ {\bf 69}, 1970 (2006); 
D.~S.~Akerib,  et al., [CDMS Collaboration],
        Phys.\ Rev.\ Lett.\ {\bf 96}, 011302 (2006).
\bibitem{john}
J.~McDonald, Phys.\ Rev.\  D {\bf 50}, 3637 (1994); 
for a recent study, see {\it e.g.}, 
H.~Sung Cheon, S.~K.~Kang and C.~S.~Kim, 
  J. Cosmol. Astropart. Phys. 05 (2008) 004.
\bibitem{ewbg-thdm2}
  S.~Kanemura, Y.~Okada and E.~Senaha,
  Phys.\ Lett.\  B {\bf 606}, 361 (2005).
 \bibitem{hhh-measurement}
  M.~Battaglia, E.~Boos and W.~M.~Yao,
  arXiv:hep-ph/0111276;  
Y.~Yasui, et al., arXiv:hep-ph/0211047.
 \bibitem{wah}
  S.~Kanemura and C.~P.~Yuan,
  Phys.\ Lett.\  B {\bf 530}, 188 (2002);
  Q.~H.~Cao, S.~Kanemura and C.~P.~Yuan,
  Phys.\ Rev.\  D {\bf 69}, 075008 (2004).
 \bibitem{zee-ph}
  S.~Kanemura, T.~Kasai, G.~L.~Lin, Y.~Okada, J.~J.~Tseng and C.~P.~Yuan,
  Phys.\ Rev.\  D {\bf 64}, 053007 (2001).
\bibitem{hagiwara}
  B.~K.~Bullock, K.~Hagiwara and A.~D.~Martin,
  Phys.\ Rev.\ Lett.\  {\bf 67}, 3055 (1991).
\end{thebibliography}
\end{document}